\documentstyle[12pt,epsf]{article}
\begin{document}
\title{\bf The Similarity Renormalization Group}
\author{S\'ergio Szpigel and Robert J. Perry
\\ \\{ \it Department of Physics}\\ { \it The Ohio State University,
Columbus, OH 43210}\\}    

\date{}
\maketitle
\abstract{Quantum field theories require a cutoff to regulate divergences that
result from local interactions, and yet physical results can not depend on the
value of this cutoff. The renormalization group employs a transformation that
changes the cutoff to isolate hamiltonians that produce cutoff-independent
eigenvalues. The similarity renormalization group is based on similarity
transformations that regulate off-diagonal matrix elements, forcing the
hamiltonian towards a band-diagonal form as the cutoff is lowered. This avoids
pathologies that plagued tradition transformations acting on hamiltonians,
making it possible to produce a well-behaved perturbative approximation of
renormalized hamiltonians in asymptotically free theories. We employ a simple
two-dimensional delta function example to illustrate this new renormalization
technique. 
}

\vfill
\eject
\vskip .25in
%

\section{Introduction}

Early attempts to combine quantum mechanics and special relativity led to the
consideration of local interactions, which are consistent with causality and
avoid signals that propagate faster than light. Local interactions lead to
divergences in perturbation theory, whose discovery caused some of the best
theorists in the world to question the foundations of quantum mechanics.
Eventual successes at fitting precise atomic experimental data led to the
universal acceptance of renormalization recipes that were acknowledged to make
little sense~\cite{schwinger}. Initially the perturbative renormalization of
QED required theorists to match perturbative expansions in powers of a bare and
physical electronic charge~\cite{dirac1}, but the bare charge clearly diverges
logarithmically in QED and the success of an expansion in powers of such a
coupling was mysterious at best~\cite{dirac3}.

The first steps towards making sense of renormalization theory were taken
in the 1950's with the invention of the perturbative renormalization group
~\cite{stueck,gellmannlow,landaupole,bogol1,bogol}, although serious
investigators found the theory was still plagued by non-convergent sums
because QED is not asymptotically free. The development of Wilson's
renormalization group formalism~\cite{wilson9,wilson10} and the discovery of
asymptotic freedom~\cite{free} allowed physicists to produce a logically
reasonable picture of renormalization in which perturbative expansions at any
high energy scale can be matched with one another, with no necessity to deal
with intermediate expansions in powers of a large parameter.

In this pedagogical article we take
advantage of the fact that the divergences in field theory result entirely from
local interactions. To understand the most important aspects of
renormalization theory requires only a background in nonrelativistic quantum
mechanics, because as has been long known the divergences of field theory are
directly encountered when one tries to impose locality on the Schr{\"o}dinger
equation. In this case the interactions we consider that are local at all
scales are delta functions and derivatives of delta functions. These
divergences can be regulated by the introduction of a cutoff, and the
artificial effects of this cutoff must be removed by renormalization. The
simplicity of the one-body Schr{\"o}dinger equation makes it possible to
renormalize the theory exactly, disentangling the effects of locality from the
complicated many-body effects and symmetry constraints encountered in
realistic field theories. There is a large literature on the
subject~\cite{zeldo}-\cite{phillips}, primarily pedagogical.

The similarity renormalization group (SRG) is a very recent development
invented by Stan G{\l}azek and Ken Wilson~\cite{wilgla1,wilgla2}, and
independently by Franz Wegner~\cite{wegner}. We do not review the applications
of this method, which are growing in number.

In the SRG, as in Wilson's original renormalization group
formalism~\cite{wilson65,wilson70}, transformations that explicitly run the
cutoff are developed. These transformations are the group elements that give
the renormalization group its name.

In his earliest work ~\cite{wilson65,wilson70} Wilson exploited a
transformation originally invented by Claude Bloch~\cite{bloch}. It uses a
cutoff on the states themselves, and as the cutoff is lowered, states are
removed from the Hilbert space. If the hamiltonian is viewed as a matrix,
these cutoffs can be seen as limiting the size of this matrix and the
transformation reduces this size, as illustrated in Fig. 1a. Wilson introduced
a rescaling operation to allow transformed hamiltonians to be compared with
initial hamiltonians, despite the fact that they act on different spaces;
however, the Bloch transformation is ill-defined and even in perturbation
theory it leads to artificial divergences. These divergences come from the
small energy differences between states retained and states removed by the
transformation, and they appear in the form of small energy denominators in
the perturbative expansion of the transformed hamiltonian. These small energy
denominator problems led Wilson to abandon the hamiltonian formulation of field
theory in favor of path integral formulations, but the virtues of the
hamiltonian formulation over the path integral formulation for many
problems remains.

\begin{figure}
\centerline{\epsffile{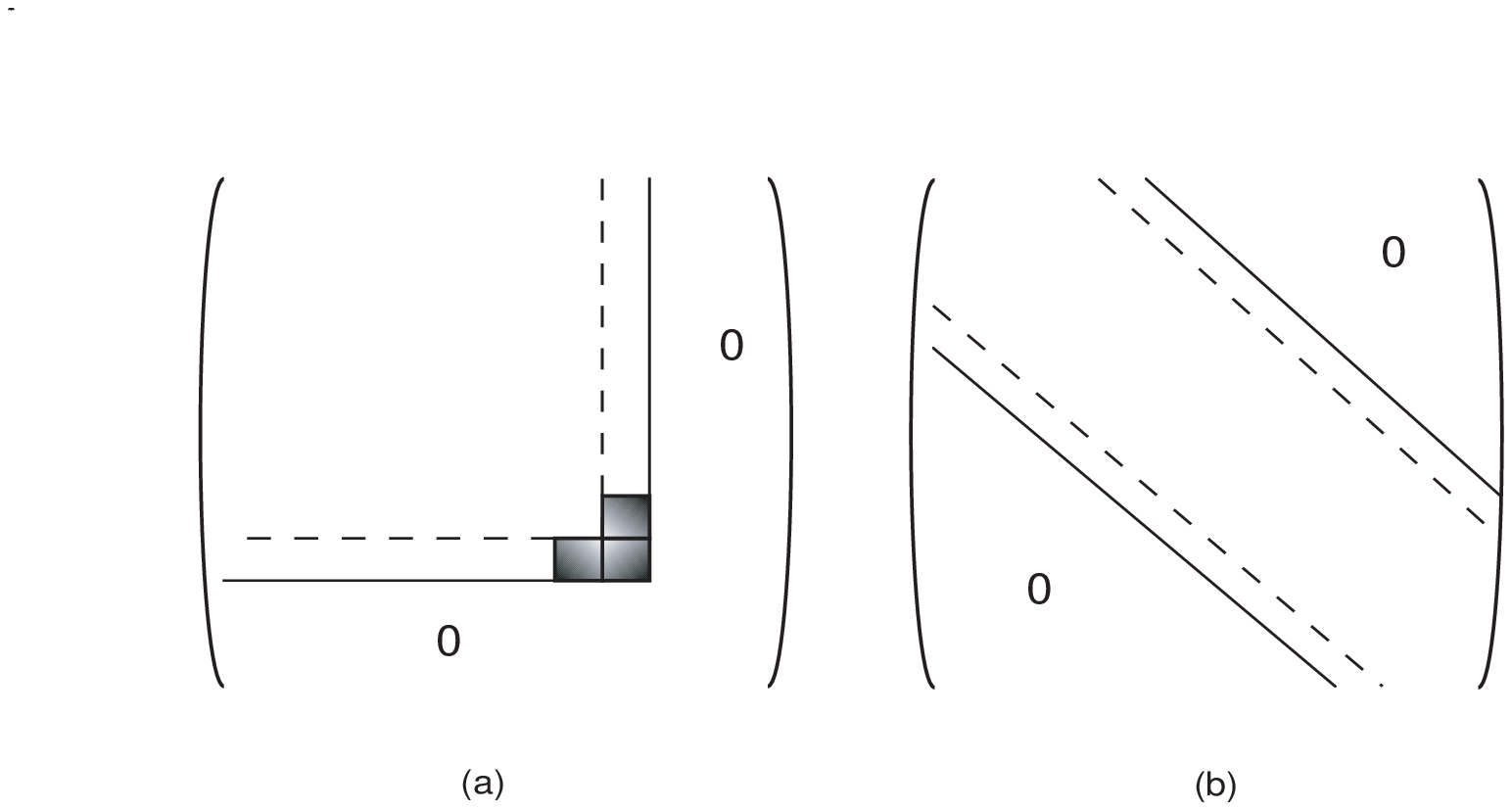}}
\vspace{0.5cm}
\caption{Two ways to run a cutoff on free energy.  In (a) a cutoff
on the magnitude of the energy is lowered from the solid to the
dashed lines, with problems resulting from the removed shaded
region. In (b) a cutoff on how far off diagonal matrix elements
appear is lowered from the dashed to the solid lines.}
\end{figure}

The breakthrough provided by the SRG is that the transformations are
typically unitary, making them well-defined, and they run a cutoff on
energy differences rather than on individual states, as illustrated in Fig. 1b.
Again viewing the hamiltonian as a large matrix, these cutoffs limit the
off-diagonal matrix elements and as they are reduced the hamiltonian is forced
towards diagonal form. The perturbative expansion for transformed hamiltonians
contains no small energy denominators, so the expansion breaks down only
when interactions become sufficiently strong, in which case perturbation
theory should fail in any case.

Although the SRG has not yet been applied to a wide range of problems, it may
be an important new tool both for attacking field theories and
non-relativistic many-body problems.

When the SRG is used with coupling coherence ~\cite{oehme,coupcoh}, which we
explain below, it allows us to construct effective theories with
the same number of free parameters as the underlying `fundamental' theory. For
the delta-function example there is one fundamental parameter, the strength
of the regulated delta-function as the cutoff is removed. In the SRG with
coupling coherence, there is only one fundamental coupling and all new
couplings are perturbative functions of the fundamental coupling that are
given by coupling coherence. It is the renormalization group flow of the added
couplings, and a boundary condition that they vanish when the fundamental
coupling is taken to zero, that fixes their dependence on the fundamental
coupling.

The examples we use in this article do not illustrate non-Gaussian fixed
points, so their scaling properties are driven by naive dimensional
analysis. However, we will see that even in these cases scaling behavior
of effective hamiltonians derived using a perturbative similarity
renormalization group can be very complicated. We will see that in the
perturbative SRG there are errors arising from the approximate
treatment of the fundamental running coupling and the approximate treatment of
the relation between this coupling and the new couplings of irrelevant
operators. 

In a realistic calculation the marginal coupling, which corresponds
to the strength of the regulated delta function, would be fit to data. In
order to clearly illustrate the logarithmic errors that result from using
the perturbative SRG equations, we approximate this marginal coupling in this
article rather than renormalizing it nonperturbatively by fitting data. The
strengths of the irrelevant operators, which correspond to derivatives of
the regulated delta function, are approximated using expansions in powers
of the approximate running coupling that are fixed by coupling coherence.
The approximate running coupling differs from the exact running coupling
by inverse powers of logarithms of the cutoff, and the error analysis for
the binding energy displays the resultant inverse logarithmic errors in
addition to power-law errors seen in all approximate renormalization group
calculations. In addition there are errors in the strengths of the irrelevant
operators resulting from using a truncated expansion in powers of the running
coupling and an approximate running coupling, both of which introduce inverse
logarithmic errors in addition to the power-law errors normally seen.

The utility of the renormalization group rests on our ability to accurately
determine and control the magnitude of errors resulting from the artificial
cutoff. For perturbative calculations this issue is not critical, but in all
field theories (and in our example) a scale is reached where the coupling
becomes large and a non-perturbative calculation must be done. The
renormalization group allows us to eliminate as much perturbative physics as
possible ({\it i.e.}, lower the cutoff as far as possible in an asymptotically
free theory), so that the essential degrees of freedom that couple
non-perturbatively can be isolated.

%

\section{Similarity Renormalization Group}

In this section we review the general formulation of the SRG developed by 
G{\l}azek and Wilson ~\cite{wilgla1,wilgla2} and a specific transformation
developed by  Wegner ~\cite{wegner}. The reader may wish to skip the general
formulation on a  first reading.

\subsection{G{\l}azek-Wilson Formulation}

Consider a system described by a hamiltonian written in the form
\begin{equation}
H=h+V \; ,
\end{equation}
where $h$ is the free hamiltonian and $V$ is an interaction.

In general, the hamiltonian can couple states of all energy scales and such  
couplings can be a source of ultraviolet divergences. The goal of the SRG is to 
obtain an effective hamiltonian in which the couplings  between high and 
low-energy states are removed, while avoiding any problems from small energy 
denominators in effective interactions. The procedure is implemented by a
unitary transformation that  generates effective interactions that reproduce
the effects of the eliminated  couplings. The effective hamiltonian  cannot
produce ultraviolet divergences at  any order in perturbation theory as long
as its matrix elements are finite. 

In our discussion we will use the basis of eigenstates of the free hamiltonian,

\begin{equation}
h |i >=\epsilon_i|i > \; .
\end{equation}
\noindent
We start by defining a bare hamiltonian, $H_\Lambda$, regulated by a very large 
cutoff $\Lambda$ (here with dimensions of energy) on the change in free energy 
at the interaction vertices,
\begin{eqnarray}
H_\Lambda&\equiv&h+V_{\Lambda} \;,\\
V_{\Lambda}&\equiv& f_{\Lambda} {\overline V}_{\Lambda}\;,\\
{\overline V}_{\Lambda}&\equiv& v+H_{\Lambda}^{ct}
\;,
\end{eqnarray}
where $f_{\Lambda}$ is a ``similarity function", ${\overline V}_{\Lambda}$ is 
defined as the reduced interaction and $H_{\Lambda}^{ct}$ are counterterms that 
must be determined through the process of renormalization in order to remove 
$\Lambda$ dependence in physical quantities.   

The similarity function $f_{\Lambda}$ regulates the hamiltonian by suppressing 
matrix elements between free states with significantly large energy difference 
and acts in the following way:
\begin{eqnarray}
< i | f_{_{\Lambda}} H_{_{\Lambda}} |j > &\equiv& \epsilon_i \; 
\delta_{ij}+f_{_{\Lambda }}(\epsilon_i-\epsilon_j) < i |{\overline 
V}_{_{\Lambda}} |j > \nonumber\\
&\equiv& \epsilon_i \; \delta_{ij}+ f_{_{\Lambda ij}} {\overline V}_{_{\Lambda 
ij}}.
\end{eqnarray}
\noindent
\noindent
Typically, the similarity function is chosen to be a smooth function 
satisfying 
\begin{eqnarray}
&&(i) f_{_{\Lambda }}(\epsilon_i-\epsilon_j) \rightarrow 1, \; {\rm when} \; 
|\epsilon_i-\epsilon_j|<< \Lambda \; ,\nonumber\\
&&(ii)f_{_{\Lambda }}(\epsilon_i-\epsilon_j)  \rightarrow 0, \; {\rm when} \; 
|\epsilon_i-\epsilon_j| >>  \Lambda \; .
\end{eqnarray}
\noindent
In several papers the similarity function has been chosen to be a step 
function.  Although  useful for doing analytic calculations, such a choice can 
lead to pathologies.

The similarity transformation  is defined to act on the bare regulated 
Hamiltonian, $H_\Lambda$, lowering the cutoff down to a scale $\lambda$:
\begin{eqnarray}
H_\lambda&\equiv& U(\lambda ,\Lambda)\; H_\Lambda \; U^\dagger(\lambda, 
\Lambda)\;.
\label{eq:star}
\end{eqnarray}
\noindent
The renormalized Hamiltonian can be written in the general form
\begin{eqnarray}
H_{\lambda} &\equiv &h+V_{\lambda} \;, \\
V_{\lambda} &\equiv& f_{\lambda}\; {\overline V}_{\lambda}\;.
\end{eqnarray}
\noindent
The transformation is unitary, so  $H_\Lambda$ and $H_\lambda$ produce the same 
spectra for observables. Also, if an exact transformation is implemented, the 
physical predictions using the renormalized Hamiltonian must be independent of 
the cutoff $\lambda$ and $ H_{\Lambda}^{ct}$ is chosen so that they also 
become independent of $\Lambda$ as $\Lambda \rightarrow \infty$.

The unitarity condition is given by:
\begin{eqnarray}
U(\lambda, \Lambda)\;  U^\dagger(\lambda, \Lambda)&\equiv&U^\dagger(\lambda, 
\Lambda)\; U(\lambda, \Lambda)  \equiv 1
\;.\label{eq:unitary}
\end{eqnarray}
\noindent
The similarity transformation $U$ can be defined in terms of an anti-hermitian 
operator $T_{\lambda}$ ($T_\lambda^\dagger=-T_\lambda$) which generates 
infinitesimal changes of the cutoff energy scale,
\begin{eqnarray}
U(\lambda ,\Lambda)&\equiv&{\cal T} \exp \left(\int_\lambda^\Lambda 
T_{\lambda^{\prime}}\; d \lambda^{\prime}\right)\; ,
\label{eq:tdef}
\end{eqnarray}
where ${\cal T}$ orders operators from left to right in order of {\it 
increasing} energy scale $\lambda^{\prime}$. Using
\begin{equation} 
T_\lambda=U(\lambda ,\Lambda)\; \frac{dU^\dagger(\lambda ,\Lambda)}{d \lambda}
  =-\frac{dU(\lambda ,\Lambda)}{d \lambda}\; U^\dagger(\lambda ,\Lambda)\;,
\end{equation}
\noindent
and the unitarity condition Eq. (\ref{eq:unitary}), we can write 
Eq. (\ref{eq:star})  in a differential form,
\begin{eqnarray}
\frac{d H_{\lambda}}{d \lambda}&=& \left [ H_\lambda,T_\lambda\right] 
\label{eq:star2}\; .
\end{eqnarray}

This is a first-order differential equation, which is solved with the boundary 
condition  $H_\lambda |_{_{\lambda \rightarrow \Lambda}} \equiv H_\Lambda$.
The  bare Hamiltonian is typically given by the canonical Hamiltonian plus 
counterterms that must be uniquely fixed to complete the renormalization.  

The operator $T_\lambda$ is defined by specifying how 
${\overline V}_\lambda$ and $h$ depend on the  cutoff scale $\lambda$. For 
simplicity in this article, we demand that $h$ is independent of $\lambda$, 
although this may not lead to an increasingly diagonal effective hamiltonian in 
all cases. We also demand that no small energy denominators can appear in the 
hamiltonian. These constraints are implemented by the conditions
\begin{eqnarray}
\frac{d h}{d \lambda}&\equiv&0\;,
\label{eq:star4}\\ \nonumber\\
\frac{d {\overline V}_\lambda}{d \lambda}&\equiv& [V_\lambda,T_\lambda]\;.
\label{eq:star3}
\end{eqnarray}

To obtain the renormalized Hamiltonian perturbatively, expand
\begin{eqnarray}
&&{\overline V}_{\lambda}=
{\overline V}_{\lambda}^{^{(1)}}
+{\overline V}_{\lambda}^{^{(2)}}+\cdots
\label{eq:yes}~,
\\
&&T_{\lambda}=
T_{\lambda}^{^{(1)}}
+T_{\lambda}^{^{(2)}}+\cdots~,\\
&&H_{_{\Lambda}}^{^{ct}} = H_{_{\Lambda}}^{^{(2),ct}}
+H_{_{\Lambda}}^{^{(3),ct}} +\cdots\;,
\end{eqnarray}
where the superscripts denote the order in the original interaction, $V$. A 
general form of these effective interactions is
\begin{eqnarray}
{\overline V}_\lambda^{^{(i)}}&=&-\sum_{j,k=1}^\infty \delta_{(j+k,i)}
\int_\lambda^\Lambda d\lambda^\prime \; [ V_{\lambda^\prime}^{^{(j)}} , 
T_{\lambda^\prime}^{^{(k)}}]
+ H_{_{\Lambda}}^{^{(i),ct}}
 \;,
\end{eqnarray}
 for $i=2,3, \cdots$, with ${\overline V}_\lambda^{^{(1)}}=v$. For instance, 
the explicit form of the second-order effective
interaction ${\overline V}_\lambda^{^{(2)}}$ is
\begin{equation}
{\overline V}_{\lambda ij}^{^{(2)}}=
\sum_{k} V_{ik} 
{V}_{kj} \left(\frac{g^{(\lambda \Lambda)}_{ikj}}{\Delta_{ik}} +
  \frac{g^{(\lambda \Lambda)}_{jki}}{\Delta_{jk}}
  \right) +H_{\Lambda ij}^{^{(2),ct}}\;,
\end{equation}
\noindent
where
\begin{eqnarray}
&&g^{(\lambda \Lambda)}_{ikj} \equiv
  \int_\lambda^\Lambda \; d \lambda^\prime \; f_{\lambda^\prime jk} 
  \; \frac{d f_{\lambda^\prime ki}}{d \lambda^\prime}\; ,
\\
&&\Delta_{ij}=\epsilon_i-\epsilon_j \; .
\end{eqnarray}

The counterterms $H_{\Lambda}^{^{(n),ct}}$ can be determined order-by-order 
using the idea of coupling-coherence ~\cite{oehme,coupcoh}. This is 
implemented by requiring the hamiltonian to reproduce itself in form under the 
similarity transformation, the only change being explicit dependence on the 
running cutoff in the operators and the implicit cutoff dependence in a finite 
number of independent running couplings. All other couplings depend on the 
cutoff only through their dependence on the independent couplings.  In general,  
we also demand  the dependent couplings to vanish when the independent 
couplings are taken to zero; i.e, the interactions are turned off. If the only 
independent coupling  in the theory is $\alpha_{\lambda}$, the renormalized 
hamiltonian can be written as an expansion in powers of this coupling:
\begin{equation}
H_{\lambda}=h+\alpha_{\lambda}{\cal O}^{(1)}+\alpha_{\lambda}^2{\cal O}^{(2)}+ 
\dots \; .
\end{equation}
\noindent
In this way, the effective hamiltonian obtained using the similarity 
transformation is completely determined by the underlying theory. The procedure 
can be extended to arbitrarily high orders, although it becomes increasingly 
complex both analytically and numerically. 

%


\subsection{Wegner Formulation}

The Wegner formulation of the SRG \cite{wegner} is defined in a very elegant 
way in terms of a flow equation analogous to the SRG Equation in the 
G{\l}azek-Wilson formalism \cite{wilgla1,wilgla2},
\begin{equation}
\frac{d H_s}{ds}=[H_s,T_s]
\;.
\end{equation}
\noindent
Here the hamiltonian $H_s=h+v_s$ evolves with a flow parameter $s$ that ranges 
from $0$ to $\infty$. The flow-parameter has dimensions $1/({\rm energy})^2$ 
and is given in terms of the similarity cutoff $\lambda$ by $s=1/\lambda^2$. 

In Wegner's  scheme the similarity transformation in defined by an explicit 
form for the generator of the similarity transformation, $T_s=[H_s,H_0]$, which  
corresponds to the choice of a gaussian similarity function with uniform width. 
In the original formulation,
Wegner advocates the inclusion of the full diagonal part of the hamiltonian at 
scale $s$ in  $H_0$. For a perturbative calculation of $H_s$, we can use
the free  hamiltonian, $H_0=h$. With this choice, the flow equation for the
hamiltonian is  given by 
\begin{equation}
\frac{d H_s}{ds}=[H_s,[H_s,h]]\; .
\end{equation}

The reduced interaction, ${\overline V}_{sij}$ (the interaction with the 
gaussian similarity function factored out) is defined by
\begin{eqnarray}
V_{sij}= f_{_{sij}}\; {\overline V}_{sij}
\;, \\
f_{_{sij}}=e^{-s\Delta_{ij}^2} \; .
\end{eqnarray}
Assuming that the free hamiltonian is independent of $s$, we obtain the flow 
equation for the reduced interaction,
\begin{equation}
\frac{d{\overline V}_{sij}}{ds}=\sum_k\left(\Delta_{ik}+\Delta_{jk}\right)
\; {\overline V}_{sik} \; {\overline V}_{skj} \; e^{-2s
\Delta_{ik}\Delta_{jk}}
\;,
\label{eq:wegnerbigone}
\end{equation}
\noindent
where we use 
$\Delta_{ij}^2-\Delta_{ik}^2-\Delta_{jk}^2=-2\Delta_{ik}\Delta_{jk}$.
We should emphasize that this is an exact equation.

To solve this equation we impose a 
boundary condition, $H_s |_{_{s \rightarrow s_0}} \equiv H_{s_0}$. 
Then, we make a perturbative expansion,
\begin{equation}
{\overline V}_{s}=
{\overline V}_{s}^{^{(1)}}
+{\overline V}_{s}^{^{(2)}}+\cdots\;,
\end{equation}
where the superscript implies the order in the bare interaction ${\overline 
V}_{s_{_{0}}}$. It is important to observe that counterterms are implicit in 
the bare interaction and can be determined in the renormalization process using 
coupling coherence.

At first order we have
\begin{equation}
\frac{d{\overline V}_{s_{ij}}^{^{(1)}}}{ds}=0\;,
\end{equation} 
which implies
\begin{equation}
{\overline V}_{sij}^{^{(1)}}={\overline V}_{s_{_{0}}ij}
\;,
\end{equation}
where $s$ is the final scale. Because of the dimensions of the flow parameter 
we have $s > s_{_{0}}$, corresponding to a smaller cutoff.  
The ``no cutoff limit" corresponds to $s_{_{0}}\longrightarrow 0$.

At second order we have
\begin{equation}
\frac{d{\overline 
V}_{sij}^{^{(2)}}}{ds}=\sum_k\left(\Delta_{ik}+\Delta_{jk}\right)
\; {\overline V}_{s_{_{0}}ik}\; {\overline V}_{s_{_{0}}kj} \; e^{-2s
\Delta_{ik}\Delta_{jk}}
\;.
\end{equation}
Integrating, we obtain
\begin{eqnarray}
{\overline V}_{sij}^{^{(2)}}&=&\frac{1}{2}\sum_k
{\overline V}_{s_{_{0}}ik}\; {\overline V}_{s_{_{0}}kj} \; 
\left(\frac{1}{\Delta_{ik}}+\frac{1}{\Delta_{jk}}\right)\times\nonumber\\
&&~~~~~~~~~~\times \left[e^{-2 s_0 
\Delta_{ik}\Delta_{jk}}-e^{-2s\Delta_{ik}\Delta_{jk}}\right]
\;.
\end{eqnarray}

By construction, the Wegner transformation is unitary and avoids small energy 
denominators. The Wegner transformation is one of the G{\l}azek-Wilson 
transformations, with the similarity function chosen to be $f_{_{\lambda 
ij}}=e^{-\Delta_{ij}^2/\lambda^2}$. 
%

%

\subsection{Strategy}

In our applications of the SRG we use Wegner's transformation. The renormalized 
hamiltonian for the non-relativistic delta-function potential in D-dimensions 
is given by 
\begin{equation}
H_{\lambda}({\bf p},{\bf p'})=p^2 \delta^{(D)}({\bf p}-{\bf p'}) + 
e^{-\frac{(p^2-p'^2)^2}{\lambda^4}}\; \left[{\bar V}_{\lambda}^{(1)}({\bf 
p},{\bf p'})+{\bar V}_{\lambda}^{(2)}({\bf p},{\bf p'})+... \right]\; ,
\label{renh}
\end{equation}
\noindent
where
\begin{eqnarray}
{\bar V}_{\lambda}^{(1)}({\bf p},{\bf 
p'})&=&-\frac{\alpha_{\lambda,i}}{(2\pi)^D} \; ,\\
{\bar V}_{\lambda}^{(2)}({\bf p},{\bf p'})&=&\alpha_{\lambda,i}^2 \; 
F^{(2)}_{s}({\bf p},{\bf p'}) \; , \\
{\bar V}_{\lambda}^{(n)}({\bf p},{\bf p'})&=&\alpha_{\lambda,i}^n \; 
F^{(n)}_{s}({\bf p},{\bf p'}) \; .
\end{eqnarray}
\noindent
Here $\lambda$ is a momentum cutoff (as opposed to the energy cutoff
discussed above) related to the flow parameter by 
$s=1/{\lambda^4}$. The index $i$ denotes the order of the calculation for the 
running coupling.

The renormalized hamiltonian can be used to compute eigenvalues and 
eigenstates. Since the hamiltonian is derived perturbatively we expect cutoff 
dependent errors in the observables. Formally, we can regroup the terms in the 
renormalized hamiltonian and write it as a momentum  expansion, and the
expansion  parameters are analytic functions of the running coupling
$\alpha_{\lambda}$.  Expanding  the operators $F^{(n)}_{s}({\bf p},{\bf p'})$
in powers of 
$p^2/{\lambda^2}$ we obtain
\begin{equation}
F^{(n)}_{s}({\bf p},{\bf p'})=z_0+z_2 \; \frac{(p^2+p'^2)}{2\lambda^2}+z_4 \; 
\frac{(p^4+ p'^2)}{4\lambda^4}+ z_4' \; \frac{p^2 p'^2}{2\lambda^4}+\cdots \; ,
\end{equation}
where the ${z_i}^{'}s$ are constants. Regrouping the terms we obtain
\begin{equation}
H_{\lambda}(p,p')=p^2 \delta^{(D)}({\bf p}-{\bf p'}) + 
e^{-\frac{(p^2-p'^2)^2}{\lambda^4}}\; 
\left[g_0(\alpha_{\lambda})+g_2(\alpha_{\lambda})\frac{( 
p^2+p'^2)}{2\lambda^2}+ \dots  \right]\; ,
\label{mexp}
\end{equation}
\noindent
where
\begin{equation}
g_i(\alpha_{\lambda})=a_i\; \alpha_{\lambda}+b_i\; 
\alpha_{\lambda}^2 + \dots \; .
\label{coupexp}
\end{equation}
\noindent

We can identify three interdependent sources of errors in the perturbative 
similarity renormalization group when the hamiltonian given by Eq.~(\ref{mexp}) 
is truncated and used to compute a physical quantity:
\vspace{0.5cm}

\noindent
a) errors introduced by the truncation of the hamiltonian at a given order in 
$p^2/\lambda^2$;
\vspace{0.3cm}

\noindent
b) errors introduced by the truncation of the hamiltonian at a given order in 
the running coupling $\alpha_{\lambda,i}$, which correspond to the use of an 
approximation for the functions $g_i$;
\vspace{0.3cm}

\noindent
c) errors introduced by the approximation for the running coupling 
$\alpha_{\lambda,i}$.
\vspace{0.5cm}

In the actual calculation using the hamiltonian given by Eq.~(\ref{renh}) 
errors of type (a) do not appear directly because we do not truncate the
operators that appear in the hamiltonian. However, errors of type (b) can be
understood as coming from approximating the couplings in front of the
operators in Eq.~(\ref{mexp}). Errors of type (c) appear in our calculations
only because we do not fit the canonical coupling to data at each scale, but
fix it at a given scale and evolve it perturbatively from that scale. The
strategy  we would use for a realistic   theory ({\it e.g.}, QED and QCD) is
the following:
\vspace{0.5cm}

\noindent
1) Obtain the renormalized hamiltonian using the similarity transformation and 
coupling-coherence, truncating the hamiltonian at a given order in powers of
$\alpha_{\lambda,i}$.
\vspace{0.3cm}

\noindent
2) Fix the coupling $\alpha_{\lambda}$ by fitting an observable ({\it e.g.}, a 
bound-state energy).
\vspace{0.3cm}

\noindent
3) Evaluate other observables ({\it e.g.}, scattering phase shifts). 
\vspace{0.5cm}

As pointed out before, the evaluation of scattering observables with the 
similarity hamiltonian with standard techniques is complicated and so in our 
examples  we focus on the bound state errors. We fix the coupling at some scale  
using a given renormalization  prescription  and use the flow-equation to 
obtain the coupling as a function of the cutoff $\lambda$ to a given order. We 
then perform a sequence of bound-state calculations with better approximations 
for the hamiltonian such that the errors in the bound-state energy are 
systematically reduced. Once the sources of errors are identified, it becomes 
relatively simple to analyze order-by-order how such errors scale with the 
cutoff $\lambda$. In principle, to completely eliminate the errors proportional 
to some power $m$ in the momentum expansion we should use the similarity 
hamiltonian with the exact running coupling (renormalized to all orders) and 
include the contributions up to ${\cal O}(p^m/{\lambda^m})$ coming from all 
effective interactions (all orders in $\alpha_{\lambda}$). Some details of
this scaling  analysis are presented later for the specific examples we work
out. We emphasize again that in a realistic calculation we would fit the
coupling  $\alpha_{\lambda}$ to an observable. This nonperturbative
renormalization  eliminates the dominant source of errors we display in SRG
calculations in this  paper. We choose to renormalize the coupling
perturbatively in this paper  because the only observable we compute is the
single bound state energy of a  delta-function potential, and fitting this
energy would prevent us from  displaying errors. 

%

%

\section{Two-Dimensional Delta-function Potential}

We now consider the case of two nonrelativistic particles in two dimensions 
interacting via an attractive  Dirac delta-function potential. The  
Schr\"{o}dinger equation for relative motion in position space (with
$\hbar=1$), can be written as:
\begin{equation}
-{\bf \nabla}_{\bf r}^2 \Psi({\bf r})-\alpha_0 \;\delta^{(2)}({\bf r}) \; 
\Psi({\bf r})=E \; \Psi({\bf r})\; .
\end{equation}
\noindent
Both the delta-function potential in two dimensions and the kinetic energy 
operator  scale as $1/r^2$, therefore, the coupling  $\alpha_0$ is 
dimensionless. As a consequence, the hamiltonian is scale invariant ({\it
i.e.},  there is no intrinsic energy scale) and we can anticipate the presence
of   logarithmic ultraviolet divergences, analogous to those appearing in QED
and QCD. The problem requires  renormalization. In this subsection we present
the standard method that  produces an exact solution analytically, using simple
regularization and  renormalization schemes \cite{jackiw}.

\subsection{Exact Solution}

We start with the Schr\"{o}dinger equation in momentum space,
\begin{equation}
p^2 \;  \Phi({\bf p})-\frac{\alpha_0}{(2\pi)^2}\; \int \; d^{2}q \;\;\Phi({\bf 
q})=E \; \Phi({\bf p}) \; ,
\label{semom2}
\end{equation}
\noindent
where $\Phi({\bf p})$ is the Fourier transform of the position space 
wave-function,
\begin{equation}
\Phi({\bf p})=\frac{1}{{2\pi}}\int \; d^{2}r  \;\; \Psi({\bf r})\;e^{-i {\bf 
p}.{\bf r}} \; .
\end{equation}

As a consequence of scale invariance, if there is any negative energy solution
to  Eq. (\ref{semom2}) then it will admit solutions for any $E<0$. This 
corresponds  to a continuum of bound states with energies extending down to
$-\infty$, so  the system is not bounded from below. By rearranging the terms
in the  Schr\"{o}dinger equation we obtain   
\begin{equation}
\Phi({\bf p})=\frac{\alpha_0}{2\pi}\; \frac{\Psi(0)}{(p^2+E_0)} \; ,
\label{2ds}
\end{equation}
\noindent
where $\Psi(0)$ is the position space wave-function at the origin and $E_0 >0$ 
is the binding energy.

To obtain the eigenvalue condition for the binding energy, we can integrate 
both sides of Eq.~(\ref{2ds}):
\begin{equation}
1=\frac{\alpha_0}{2\pi}\;  \int_{0}^{\infty} \; dp \; p \; \frac{1}{(p^2+E_0)} 
\; .
\label{bsi2}
\end{equation}
\noindent
The integral on the r.h.s. diverges logarithmically, so the problem is 
ill-defined. 

The conventional way to deal with this problem is renormalization. First, we 
regulate the integral with a momentum cutoff, obtaining
\begin{equation}
1=\frac{\alpha_0}{2\pi}\;  \int_{0}^{\Lambda} \; dp \; p \; 
\frac{1}{(p^2+E_0)}=\frac{\alpha_0}{4\pi}\; {\rm 
ln}\left(1+\frac{\Lambda^2}{E_0}\right) \; ,
\end{equation}
\noindent
so that 
\begin{equation}
E_0=  \frac{\Lambda^2}{e^{-\frac{4\pi}{\alpha_0}}-1} \; .
\end{equation}

Clearly, if the coupling $\alpha_0$ is fixed then $E_0 \rightarrow \infty$ as 
$\Lambda \rightarrow \infty$. In order to eliminate the divergence and produce 
a finite, well-defined bound state we can renormalize the theory by demanding 
that the coupling runs with the cutoff $\Lambda$ in such a way that the binding 
energy remains fixed as the cutoff is removed:
\begin{equation}
\alpha_0 \rightarrow \alpha_{\Lambda}=\frac{4\pi}{{\rm 
ln}\left(1+\frac{\Lambda^2}{E_0}\right)} \; .
\end{equation}

The dimensionless renormalized running coupling $\alpha_{\Lambda}$  that 
characterizes the strength of the interaction is therefore replaced by a new 
(dimensionful) parameter $E_0 >0$, the binding energy of the system. This is a 
simple example of dimensional transmutation \cite{coleman}: even though the
original ``bare''  hamiltonian is scale invariant, the renormalization
procedure leads to a scale  that characterizes the physical observables. Note
that $E_0$ can be chosen  arbitrarily, fixing the energy scale of the
underlying (renormalized) theory. It  is also interesting to note that the
renormalized running coupling 
$\alpha_{\Lambda}$ vanishes as $\Lambda \rightarrow \infty$ and so the theory 
is asymptotically free.

This renormalized hamiltonian can be used to compute other observables. The 
usual prescription for the calculations is to obtain the solutions with the 
cutoff in place and then take the limit as the momentum cutoff is removed to 
$\infty$. If an exact calculation can be implemented, the final results should 
be independent of the regularization and renormalization schemes. As an 
example, we calculate the scattering wave function,
\begin{equation}
\Phi_{k}({\bf p})=\delta^{(2)}({\bf p}-{\bf k})+\frac{\alpha_{\Lambda}}{2 
\pi}\; \frac{\Psi(0)}{(p^2-k^2-i \; \epsilon)} \; ,
\end{equation}
\noindent
where $k=\sqrt{E}$.
\noindent
Integrating both sides over ${\bf p}$ with a cutoff $\Lambda$ in place, we 
obtain
\begin{equation}
\Psi(0)=\frac{1}{2\pi} \; \left[1-\frac{\alpha_{\Lambda}}{4\pi}\; {\rm 
ln}\left(1+\frac{\Lambda^2}{-k^2-i\epsilon}\right)\right]^{-1} \; ;
\end{equation}
\noindent
thus,
\begin{equation}
\alpha_{\Lambda} \; \Psi(0)=\frac{1}{2\pi} \; \left[\frac{1}{4\pi}\;{\rm 
ln}\left(1+\frac{\Lambda^2}{E_0}\right)-\frac{1}{4\pi}\; {\rm 
ln}\left(1+\frac{\Lambda^2}{-k^2-i\epsilon}\right)\right]^{-1} \; .
\end{equation}
\noindent
In the limit $\Lambda \rightarrow \infty$ we obtain:
\begin{equation}
\alpha_{\Lambda} \; \Psi(0)=\frac{2}{{\rm ln}\left(\frac{k^2}{E_0}\right)-i \; 
\pi} \; .
\end{equation}
\noindent
The resulting scattering wave function is then given by
\begin{equation}
\Phi_{k}({\bf p})=\delta^{(2)}({\bf p}-{\bf k})+\frac{1}{2 \pi}\; 
\frac{2}{(p^2-k^2-i\epsilon)}\left[{\rm ln}\left(\frac{k^2}{E_0}\right)-i \; 
\pi\right]^{-1} \; .
\end{equation}

It is important to note that only S-wave scattering occurs, corresponding to 
zero angular momentum states. For the higher waves  the centrifugal barrier 
completely screens the delta-function potential and the non-zero angular 
momentum scattering states are free states. 

The same prescription  can be used to  evaluate the T-matrix or the K-matrix.
For the T-matrix,  the Lippmann-Schwinger equation with the renormalized 
potential is given by:
\begin{equation}
T({\bf p},{\bf p'};k)=V({\bf p},{\bf p'})+\int \; d^{2}q \; \; \frac{V({\bf 
p},{\bf q})}{k^2-q^2+i \epsilon} \; T({\bf q},{\bf p'};k) \; .
\end{equation}
\noindent
Since only S-wave scattering takes place we can integrate out the angular 
variable, obtaining 
\begin{equation}
T^{(\rm l=0)}( p,p';k)=V^{(\rm l=0)}(p,p')+\int_{0}^{\Lambda} \; dq \; q \;  
\frac{V^{(\rm l=0)}(p,q)}{k^2-q^2+i \epsilon} \; T^{(\rm l=0)}(q,p';k) \; ,
\end{equation}
\noindent
where
\begin{equation}
V^{(\rm l=0)}(p,p')=-\frac{\alpha_{\Lambda}}{2\pi} \; .
\end{equation}

The Lippmann-Schwinger equation for the ``on-shell'' T-matrix is given by:
\begin{eqnarray}
T^{(\rm l=0)}(k)=-\frac{\alpha_{\Lambda}}{2\pi}-\frac{\alpha_{\Lambda}}{2\pi}\; 
T^{(\rm l=0)}(k) \; \int_{0}^{\Lambda} \; dq \; q \;  \frac{1}{k^2-q^2+i 
\epsilon}.
\end{eqnarray}
\noindent
Solving this equation and taking the limit $\Lambda \rightarrow \infty$, we 
obtain the exact ``on-shell'' T-matrix:
\begin{equation}
T_{0}(k)= -\frac{2}{{\rm ln}\left(\frac{k^2}{E_0}\right)-i \; \pi}\; .
\end{equation}
\noindent
Here and in what follows we drop the superscript and use the subscript $0$ to 
denote the exact quantities.

In the same way, the S-wave Lippmann-Schwinger equation for the K-matrix is 
given by
\begin{equation}
K(p,p';k)=V(p,p')+{\cal P}\int_{0}^{\Lambda} \; dq \; q \; 
\frac{V(p,q)}{k^2-q^2} \; K(q,p';k) \; ,
\end{equation}
\noindent
and the exact ``on-shell'' K-matrix is given by
\begin{equation}
K_0(k)= -\frac{2}{{\rm ln}\left(\frac{k^2}{E_0}\right)}\; .
\end{equation}

The ``on-shell'' K-matrix and T-matrix  are related by
\begin{equation}
K_0(k)=\frac{T_0(k)}{1-\frac{i\pi}{2}T_0(k)} \; .
\end{equation}
\noindent
Using either 
\begin{equation}
k \; {\rm cot}\;\delta_{0}(k)-i k=-\frac{2 \; k}{\pi}\frac{1}{T_0(k)} \; ,
\end{equation}
\noindent
or
\begin{equation}
k \;{\rm cot}\; \delta_0(k)=-\frac{2 \; k}{\pi}\frac{1}{K_0(k)} \; ,
\end{equation}
\noindent
we can obtain the exact phase-shifts:
\begin{equation}
{\rm cot} \; \delta_{0}=\frac{1}{\pi}\; {\rm ln}\left(\frac{k^2}{E_0}\right)\; 
.
\end{equation}
%

%

%

\subsection{Similarity Renormalization Group Approach}

In the two-dimensional case the canonical hamiltonian in momentum space with a 
delta-function potential can be written as 
\begin{equation}
H({\bf p},{\bf p'})=h({\bf p},{\bf p'})+V({\bf p},{\bf p'}) \; , 
\end{equation}
\noindent
where $h({\bf p},{\bf p'})=p^2 \delta^{(2)}({\bf p}-{\bf p'})$ corresponds to 
the free hamiltonian and $V({\bf p},{\bf p'})=-{\alpha_0}/(2\pi)^2$ corresponds 
to the Fourier transform of the delta-function potential.

Integrating out the angular variable, the flow equation obtained with Wegner's 
transformation in terms of matrix elements in the basis of free states is given 
by 
\begin{equation}
\frac{dV_s(p,p')}{ds}=-(p^2-p'^2)^2 \; V_{s}(p,p')-\int_{0}^{\infty}dk \; k \; 
(2 k^2-p^2-p'^2)\; V_{s}(p,k)\; V_{s}(k,p') \; .
\end{equation}
\noindent
In principle, we can set the boundary condition at $s=0$ (no cutoff), i.e, 
\begin{equation}
H_{s=0}(p,p')=H(p,p')=p^2 \delta^{(1)}(p-p')-\frac{\alpha_0}{2\pi} \; .
\end{equation}
\noindent
However, the hamiltonian with no cutoff produces logarithmic divergences, 
requiring renormalization. As we will see, the boundary condition must be 
imposed at some other point, leading to dimensional
transmutation~\cite{coleman}.
\noindent
The reduced interaction ${\bar V}_{s}(p,p')$ is defined such that
\begin{equation}
V_{s}(p,p')=e^{-s(p^2-p'^2)^2}\; {\bar V}_{s}(p,p') \; .
\end{equation}
\noindent
Assuming that $h$ is cutoff independent we obtain the flow equation for the 
reduced interaction, 
\begin{eqnarray}
\frac{d{\bar V}_{s}}{ds}=-e^{-2s\; p^2  p'^2}\; \int_{0}^{\infty}&&dk \; k \; 
(2 k^2-p^2-p'^2)\; e^{-2s[k^4-k^2(p^2+p'^2)]}\nonumber\\
&&\times {\bar V}_{s}(p,k)\; {\bar V}_{s}(k,p') \; .
\label{f2}
\end{eqnarray}

This equation is solved using  a perturbative expansion, starting with
\begin{equation}
{\bar V}^{(1)}_{s}(p,p')=-\frac{{\alpha}_s}{2\pi} \; .
\end{equation}
\noindent
We assume a coupling-coherent solution in the form of an expansion in powers of 
$\alpha_s/2\pi$, satisfying the constraint that the operators 
$F^{(n)}_{s}(p,p')$ vanish when $p=p'=0$,
\begin{equation}
{\bar 
V}_{s}(p,p')=-\frac{\alpha_s}{2\pi}+\sum_{n=2}^{\infty}\left(\frac{\alpha_{s}}{
2\pi}\right)^{n}\; F^{(n)}_{s}(p,p') \; .
\label{coc2}
\end{equation}
\noindent
Note that the expansion parameter is $\alpha_s/2\pi$.

Using  the solution  Eq. (\ref{coc2}) in Eq. (\ref{f2}) we obtain
\begin{eqnarray}
\frac{d{\bar V}_{s}}{ds}&=&-\frac{1}{(2\pi)^2}\; 
\frac{d{\alpha}_s}{ds}+\sum_{n=2}^{\infty}\frac{1}{(2\pi)^n}\left[n \; 
\alpha_{s}^{n-1}\; \frac{d{\alpha}_s}{ds}\; F^{(n)}_{s}(p,p')+\alpha_{s}^{n}\; 
\frac{dF^{(n)}_{s}(p,p')}{ds}\right]\nonumber\\
&=&\int_{0}^{\infty}dk \; k \; (2 k^2-p^2-p'^2)\; e^{-2s[p^2 
p'^2+k^4-k^2(p^2+p'^2)]}\nonumber\\
&\times&\left[-\frac{\alpha_s}{2\pi}+\sum_{n=2}^{\infty}\left(\frac{\alpha_{s}}
{2\pi}\right)^{n}\; 
F^{(n)}_{s}(p,k)\right]\left[-\frac{\alpha_s}{2\pi}+\sum_{m=2}^{\infty}
\left(\frac{\alpha_{s}}{2\pi}\right)^{m}\;
F^{(m)}_{s}(k,p')\right] \; .
\end{eqnarray}
\noindent
This equation is solved iteratively order-by-order in $\alpha_s/2\pi$. Again, 
if $\alpha_s/2\pi$ is small the operator ${\bar V}^{(1)}_{s}(p,p')$ can be 
identified as the dominant term in the expansion of ${\bar V}_{s}(p,p')$ in 
powers of $p$ and $p'$. In the $D=2$ case this operator corresponds to a 
marginal operator (since the coupling is dimensionless and there is no implicit 
mass scale). The higher-order terms correspond to irrelevant operators. 

At second-order we have
\begin{equation}
-\frac{1}{2 \pi}\; \frac{d{\alpha}_s}{ds}+\frac{1}{(2\pi)^2}\alpha_{s}^{2}\; 
\frac{dF^{(2)}_{s}(p,p')}{ds}=- \alpha_s^2 \; I_{s}^{(2)}(p,p') \; ,
\end{equation}
\noindent
where
\begin{eqnarray}
I_{s}^{(2)}(p,p')&=&\frac{1}{(2\pi)^2}\; \int_{0}^{\infty}dk \; k \; (2 
k^2-p^2-p'^2)\; e^{-2s[p^2 p'^2+k^4-k^2(p^2+p'^2)]}\nonumber\\
&=&\frac{1}{(2\pi)^2}\; \frac{e^{-2s \; p^2 p'^2}}{4s} \; .
\end{eqnarray}
\noindent
The equation for $\alpha_s$ is obtained by taking the limit $(p,p') \rightarrow 
0$, 
\begin{equation}
\frac{1}{2 \pi}\; \frac{d{\alpha}_s}{ds}= \alpha_s^2 \; I_{s}^{(2)}(0,0) \; ,
\label{alp22}
\end{equation}
\noindent
where
\begin{equation}
I_{s}^{(2)}(0,0)=\frac{1}{(2\pi)^2}\; \frac{1}{4s} \; .
\end{equation}
\noindent
Integrating Eq. (\ref{alp22}) from $s_0$ to $s$,
\begin{eqnarray}
&&\alpha_{s,2}=\frac{\alpha_{s_0}}{1- \frac{\alpha_{s_0}}{8 \pi}\;{\rm 
ln}\left(\frac{s}{s_0}\right)} \; .
\label{ralp22}
\end{eqnarray}
\noindent
In terms of the cutoff $\lambda$ we obtain
\begin{equation}
\alpha_{\lambda,2}=\frac{\alpha_{\lambda_0}}{1+\frac{\alpha_{\lambda_0}}{2 
\pi}\;{\rm ln}\left(\frac{\lambda}{\lambda_0}\right)} \; .
\end{equation}
\noindent
In principle,  knowing the value of $\alpha_{s_0}$ for a given $s_0$ we can 
determine the running coupling $\alpha_s$ for any $s$.  Since we cannot choose 
$s_0=0$ ($\lambda_0=\infty$), to use Eq.~(\ref{ralp22}) we 
must specify a renormalization prescription that allows us to fix the coupling 
at some finite non-zero value of $s_0$. We discuss this issue in detail later 
in this subsection.

The equation for $F^{(2)}_{s}(p,p')$ is given by 
\begin{equation}
\frac{1}{(2\pi)^2}\frac{dF^{(2)}_{s}(p,p')}{ds}=I_{s}^{(2)}(0,0)- 
I_{s}^{(2)}(p,p') \; .
\end{equation}
\noindent
Integrating from $s_0$ to $s$ we obtain
\begin{eqnarray}
F^{(2)}_{s}(p,p')&=& \int_{s_0}^{s}ds' \; \frac{\left[1-e^{-2s' \; p^2 
p'^2}\right]}{4s'}\nonumber\\
&=& \frac{1}{4}\left[{\rm ln}\left(\frac{s}{s_0}\right)-{\rm Ei}(-2s \; p^2 \; 
p'^2)+{\rm Ei}(-2s_0 \; p^2 \; p'^2)\right]\; .
\end{eqnarray}
\noindent
Insisting that $F^{(2)}_{s}(p,p')=0$ when $p,p'=0$ we obtain
\begin{eqnarray}
F^{(2)}_{s}(p,p')&=& \frac{1}{4}\left[\gamma+{\rm ln}(2s \; p^2 \; p'^2)-{\rm 
Ei}(-2s \; p^2 \; p'^2)\right] \; .
\end{eqnarray}

At third-order we have
\begin{eqnarray}
-\frac{1}{2 \pi}\; \frac{d{\alpha}_s}{ds}+\frac{1}{(2\pi)^2}\; \alpha_{s}^{2}\; 
\frac{dF^{(2)}_{s}(p,p')}{ds}&+&\frac{2}{(2\pi)^2}\; \alpha_{s} \; 
\frac{d\alpha_{s}}{ds}\; F^{(2)}_{s}(p,p')+\frac{1}{(2\pi)^3}\; \alpha_{s}^{3} 
\; \frac{dF^{(3)}_{s}(p,p')}{ds}\nonumber\\
&=&- \alpha_s^2 \; I_{s}^{(2)}(p,p')+ \alpha_s^3 \; I_{s}^{(3)}(p,p') \; ,
\end{eqnarray}
\noindent
where
\begin{eqnarray}
I_{s}^{(3)}(p,p')&=&\frac{1}{(2\pi)^3}\;\int_{0}^{\infty}dk \; k \; (2 
k^2-p^2-p'^2)\; e^{-2s[p^2 p'^2+k^4-k^2(p^2+p'^2)]} \nonumber\\
&\times& \left[F^{(2)}_{s}(p,k)+F^{(2)}_{s}(k,p')\right] \; .
\end{eqnarray}
\noindent
In the limit $p,p' \rightarrow 0$ we obtain:
\begin{equation}
\frac{1}{2 \pi}\; \frac{d{\alpha}_s}{ds}= \alpha_s^2 \; I_{s}^{(2)}(0,0)- 
\alpha_s^3 \; I_{s}^{(3)}(0,0) \; ,
\label{alp32}
\end{equation}
\noindent
where
\begin{eqnarray}
I_{s}^{(3)}(0,0)=\frac{1}{(2\pi)^3}\; \int_{0}^{\infty}dk  \; 2 k^3 e^{-2s 
k^4}\; \left[F^{(2)}_{s}(0,k)+F^{(2)}_{s}(k,0)\right] \; .
\end{eqnarray}
\noindent
Since $ F^{(2)}_{s}(k,0)= F^{(2)}_{s}(0,k)=0$, the term proportional to 
$\alpha_s^3$ in Eq. (~\ref{alp32}) is zero.

The equation for $F^{(3)}_{s}(p,p')$ is given by 
\begin{equation}
\frac{1}{(2\pi)^3}\frac{dF^{(3)}_{s}(p,p')}{ds}=-\frac{1}{\pi} \; 
I_{s}^{(2)}(0,0)\; F^{(2)}_{s}(p,p')+ I_{s}^{(3)}(p,p') \; .
\end{equation}
\noindent
To obtain $F^{(3)}_{s}(p,p')$ the integrals in $k$ and $s$ must be evaluated 
numerically.

At fourth-order we obtain
\begin{eqnarray}
-\frac{1}{2 \pi}\; \frac{d{\alpha}_s}{ds}&+&\frac{1}{(2\pi)^2} \; 
\alpha_{s}^{2}\; \frac{dF^{(2)}_{s}(p,p')}{ds}
+\frac{2}{(2\pi)^2} \; \alpha_{s} \; \frac{d\alpha_{s}}{ds}\; 
F^{(2)}_{s}(p,p')\nonumber\\
&+&\frac{1}{(2\pi)^3}\; \alpha_{s}^{3} \; 
\frac{dF^{(3)}_{s}(p,p')}{ds}+\frac{3}{(2\pi)^3}\;\alpha_{s}^{2}\; 
\frac{d{\alpha}_s}{ds} \; F^{(3)}_{s}(p,p')
+\frac{1}{(2\pi)^4}\alpha_{s}^{4} \; \frac{dF^{(4)}_{s}(p,p')}{ds}\nonumber\\
&=&- \alpha_s^2 \; I_{s}^{(2)}(p,p')+ \alpha_s^3 \; I_{s}^{(3)}(p,p') + 
\alpha_s^4 \; I_{s}^{(4)}(p,p')  \; ,
\end{eqnarray}
\noindent
where
\begin{eqnarray}
I_{s}^{(4)}(p,p')&=&\frac{1}{(2\pi)^4}\;\int_{0}^{\infty}dk \; k \; (2 
k^2-p^2-p'^2)\; e^{-2s[p^2 p'^2+k^4-k^2(p^2+p'^2)]} \nonumber\\
&\times& \left[F^{(3)}_{s}(p,k)+F^{(3)}_{s}(k,p')+ F^{(2)}_{s}(p,k) 
F^{(2)}_{s}(k,p')\right] \; .
\end{eqnarray}
In the limit $p,p' \rightarrow 0$ we obtain:
\begin{equation}
\frac{1}{2 \pi}\; \frac{d{\alpha}_s}{ds}= \alpha_s^2 \; I_{s}^{(2)}(0,0) - 
\alpha_s^4 \; I_{s}^{(4)}(0,0) \; ,
\label{alp42}
\end{equation}
\noindent
where
\begin{eqnarray}
I_{s}^{(4)}(0,0)=\frac{1}{2\pi}\; \int_{0}^{\infty}dk \; k \; 2 k^2 e^{-2s 
k^4}\; \left[F^{(3)}_{s}(0,k)+F^{(3)}_{s}(k,0)\right] \; .
\end{eqnarray}
\noindent
For dimensional reasons Eq. (\ref{alp42}) takes the form
\begin{equation}
\frac{d{\alpha}_s}{ds}= \frac{B_2}{s} \; \alpha_s^2 - \frac{B_4}{s}\;  
\alpha_s^4 \; ,
\label{alp34}
\end{equation}
\noindent
where $B_2=\frac{1}{8\pi} $ and $B_4$ can be obtained by evaluating 
$I_{s}^{(4)}(0,0)$ (numerically) for $s=1$. 
\noindent
In terms of the cutoff $\lambda$ we obtain:
\begin{equation}
\frac{d{\alpha}_{\lambda}}{d\lambda}= \frac{1}{\lambda^{4}} \; B_2 \; 
\alpha_{\lambda}^{2} - \frac{1}{\lambda^4}\; B_4 \; \alpha_{\lambda}^{4} \; .
\end{equation}
\noindent
Integration of Eq. (\ref{alp42}) leads to a transcendental equation which is 
solved numerically in order to obtain the running coupling $\alpha_{s,4}$.

Qualitatively, the errors are expected to be a combination of inverse powers of 
$\lambda$ and powers (or inverse powers) 
of ${\rm ln}(\lambda)$ coming from the perturbative expansion in powers of 
$\alpha_{\lambda}$ for the coefficients of the irrelevant operators and the 
perturbative approximation for $\alpha_{\lambda}$. 

As pointed out above, to completely determine the renormalized hamiltonian at
a  given  order we need to specify the coupling at some scale $\lambda_0$. The
simplest  way is to choose a  value for the `exact' $\alpha_{\lambda_0}$.
Formally, this fixes the underlying  theory; {\it i.e.},  if we had the exact
hamiltonian (to all orders) we could obtain the exact values for all 
observables. However, since the hamiltonian is derived perturbatively and must 
be truncated  at some order in practical calculations, we can only obtain
approximate cutoff-dependent results for the observables. Moreover, in this
case the errors cannot be directly  evaluated, since the  exact values remain
unknown. As an example, we calculate the bound-state energy   choosing 
$\alpha_{\lambda_0}=1.45$ at $\lambda_0=100$. In Fig. 2 we show the 
binding-energy as a  function of the cutoff $\lambda$ using the following
approximations for the  interaction:
\vspace{0.5cm}

\noindent
(a) marginal operator with coupling ($\alpha_{\lambda_0}$),
\begin{equation}
V_{\lambda}(p,p')=-\frac{\alpha_{\lambda_0}}{2\pi}e^{-\frac{(p^2-p'^2)^2}
{\lambda^4}} \; ;
\end{equation}

\noindent
(b) marginal operator with running coupling renormalized to second-order 
($\alpha_{\lambda,2}$),
\begin{equation}
V_{\lambda}(p,p')=-\frac{\alpha_{\lambda,2}}{2\pi}\; 
e^{-\frac{(p^2-p'^2)^2}{\lambda^4}} \; ;
\end{equation}

\noindent
(c) marginal operator plus second-order irrelevant operator with running 
coupling renormalized to second-order ($\alpha_{\lambda,2}, 
F^{(2)}_{\lambda}$),
\begin{equation}
V_{\lambda}(p,p')=\left[-\frac{\alpha_{\lambda,2}}{2\pi}+\left(\frac{\alpha_{
\lambda,2}}{2\pi}\right)^2  \; F^{(2)}_{\lambda}(p,p')\right]\;  
e^{-\frac{(p^2-p'^2)^2}{\lambda^4}} \; ;
\end{equation}

\noindent
(d) marginal operator with running coupling renormalized to fourth-order  
($\alpha_{\lambda,4}$),
\begin{equation}
V_{\lambda}(p,p')=-\frac{\alpha_{\lambda,4}}{2\pi}\; 
e^{-\frac{(p^2-p'^2)^2}{\lambda^4}} \; ;
\end{equation}

\noindent
(e) marginal operator plus second-order irrelevant operator with running 
coupling renormalized to fourth-order $\alpha_{\lambda,4}, F^{(2)}_{\lambda}$),

\begin{equation}
V_{\lambda}(p,p')=\left[-\frac{\alpha_{\lambda,4}}{2\pi}+\left(\frac{\alpha_{
\lambda,4}}{2\pi}\right)^2  \; F^{(2)}_{\lambda}(p,p')\right] \; 
e^{-\frac{(p^2-p'^2)^2}{\lambda^4}} \; .
\end{equation}
\vspace{0.5cm}

\begin{figure}
\centerline{\epsffile{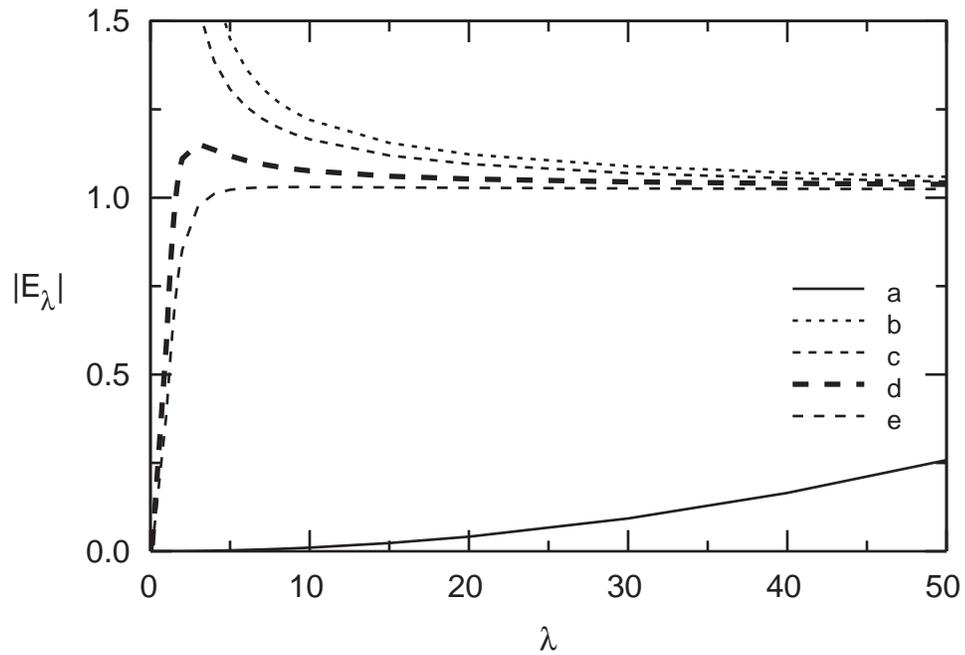}}
\vspace{0.5cm}
\caption{The binding energy for the two-dimensional delta-function potential 
with various approximations for the SRG hamiltonian. The exact theory is fixed 
by choosing $\alpha_0=1.45$ at $\lambda_0=100$.}
\end{figure}

We see that as the approximation is improved the cutoff dependence is reduced. 
As $\lambda \rightarrow \infty$ all curves should approach the same 
binding-energy, which corresponds to the exact value, and as $\lambda$ becomes 
small the perturbative approximation breaks down. 

A similar prescription is to find  $\alpha_{\lambda_0}$ at $\lambda_0$ that 
produces a given binding-energy, $E_0$. Since the fitting is implemented using 
a truncated hamiltonian, this $\alpha_{\lambda_0}$ is an approximation that 
becomes more accurate if we use a larger $\lambda_0$ and/or include higher 
order operators. Although in this case we can evaluate the errors, the scaling 
analysis becomes complicated as $\lambda \rightarrow \lambda_0$ because at this 
point we force the energy to be the exact value and so the error is zero. 

As an example, we calculate the bound-state energy when the coupling is fixed
at  $\lambda_0=100$ to give $E_0=1$. In Fig. 3 we show the `errors' in the
binding  energy using the same approximations listed above for the potential.
As  expected, all error lines drop abruptly to zero when $\lambda \rightarrow 
\lambda_0$, where the running coupling is chosen to fit what we define to be
the exact binding  energy. Away from this point we can analyze the errors.
With the hamiltonian  (a) (unrenormalized) we obtain a dominant error that
scales like ${\rm  ln}\left(\lambda_0/\lambda \right)$, corresponding to the
leading order error.  With the renormalized hamiltonian (b) the dominant
errors scale like $[{\rm  ln}\left(\lambda_0/\lambda \right)]^{-2}$, indicating
the elimination of the  leading order logarithmic errors. With the hamiltonian
(c) there is a small  shift, but no significant change in the error scaling.
The added irrelevant  operator may remove errors of order
$(\lambda_0/\lambda)^2$, but these are  smaller than the remaining 
$[{\rm ln}\left(\lambda_0/\lambda \right)]^{-2}$  errors. With hamiltonians (d)
and (e) in the range of intermediate cutoffs  ($E_0 << \lambda^2 <<
\lambda_0^2$) there is also only a shift in the errors.  The dips in (d) and
(e) correspond to values of $\lambda$ where the binding  energy equals the
exact value.

\begin{figure}
\centerline{\epsffile{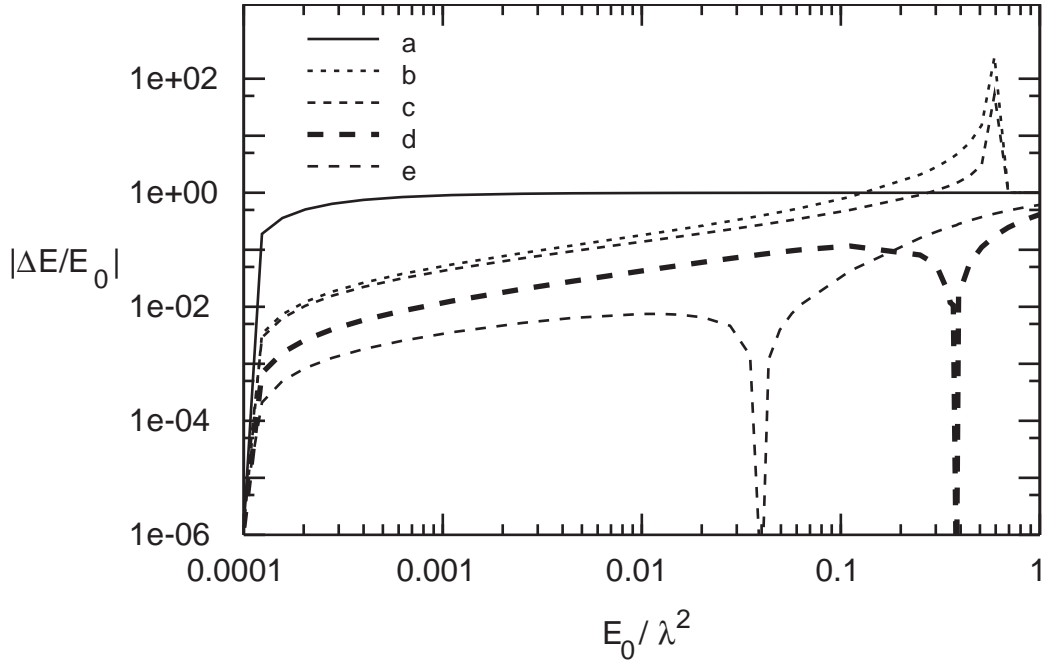}}
\vspace{0.5cm}
\caption{The SRG errors in the binding energy for the two-dimensional 
delta-function potential using various approximations for the SRG hamiltonian. 
The exact theory is fixed by choosing $E_0=1$ at $\lambda_0=100$.}
\end{figure}

This behavior is a perturbative artifact that 
can be understood in the following way. Consider the Schr\"{o}dinger equation 
with potential (d). Rescaling the momenta $p \rightarrow \lambda {\tilde p}$ we 
obtain
\begin{equation}
{\tilde p}^2 {\tilde \Phi}({\tilde p})-\frac{\alpha_{\lambda,4}}{2\pi}\; 
\int_{0}^{\infty} \; d{\tilde q} \; {\tilde q} \; e^{({\tilde p}^2-{\tilde 
q}^2)}\;{\tilde \Phi}({\tilde q})=\frac{E_{\lambda}}{\lambda^2}\; {\tilde 
\Phi}({\tilde p}) \; .
\end{equation}
\noindent
and
\begin{eqnarray}
E_{\lambda}=\lambda^2 \; \frac{\left[\int_{0}^{\infty}\; d{\tilde p} \; {\tilde 
p}\; \left({\tilde p}^2\;|{\tilde \Phi}({\tilde 
p})|^2\right)-\frac{\alpha_{\lambda,4}}{2\pi}\; \int_{0}^{\infty}\; d{\tilde p} 
\; {\tilde p}\; \int_{0}^{\infty} \; d{\tilde q} \; {\tilde q} \; e^{({\tilde 
p}^2-{\tilde q}^2)} \;{\tilde \Phi}({\tilde p})\;{\tilde \Phi}({\tilde 
q})\right]}{\int_{0}^{\infty}\; d{\tilde p} \; {\tilde p}\;|{\tilde 
\Phi}({\tilde p})|^2}\; .
\end{eqnarray}
As shown in Fig. 4, the coupling renormalized to fourth-order, 
$\alpha_{\lambda,4}$, approximately freezes for small $\lambda$ and as a 
consequence the bound-state energy scales like $E_{\lambda}\simeq \lambda^2 
\times {\rm constant}$ eventually becoming equal to the exact value and then 
deviating again. With the hamiltonian (e) the behavior is similar, with the dip 
occurring at a different value of $\lambda$ because of the irrelevant operator. 
For small values of $\lambda$ the lines converge, 
indicating the breakdown of the perturbative expansion. 

\begin{figure}
\centerline{\epsffile{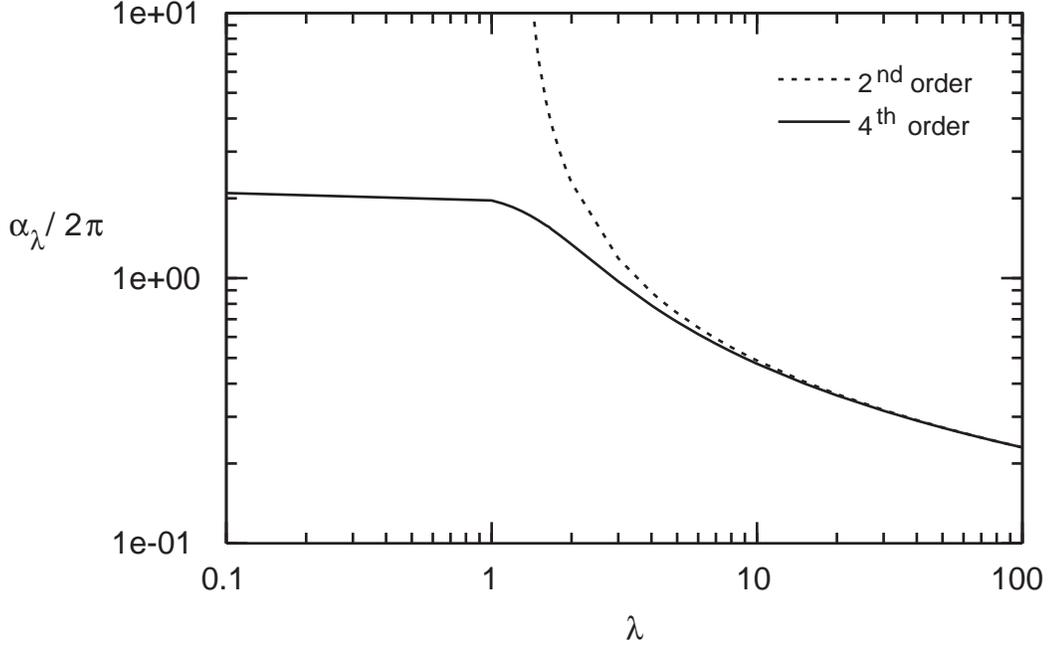}}
\vspace{0.5cm}
\caption{The SRG running coupling for the two-dimensional delta-function 
potential obtained with $\alpha_{\lambda_0}$ at $\lambda_0=100$ fixed to fit 
$E_0=1$.}
\end{figure}

An alternative prescription is to use the potential derived in subsection 5.1 
as the starting point for the similarity transformation. We introduce a large  
momentum cutoff $\Lambda$, define
\begin{equation}
\alpha_{s_0=0}=\alpha_{\Lambda}=\frac{4\pi}{{\rm 
ln}\left(1+\frac{\Lambda^2}{E_0}\right)} \; ,
\end{equation}   
\noindent
and set all irrelevant operators to zero at $s_0=0$. Note that the coupling 
$\alpha_{\lambda_0}$ is fixed at $\lambda_0=\infty$ by fitting the exact 
binding energy. With this definition the similarity hamiltonian with no
similarity cutoff  becomes well-defined and  we can set all of the similarity
transformation  boundary conditions at $s=0$. The previous derivation remains
essentially the same. The only  modification is that all integrals over
momentum are cut off ($p \le  
\Lambda$).

At second-order we have
\begin{equation}
-\frac{1}{2 \pi}\; 
\frac{d{\alpha}_{s,\Lambda}}{ds}+\frac{1}{(2\pi)^2}\alpha_{s,\Lambda}^{2}\; 
\frac{dF^{(2)}_{s,\Lambda}(p,p')}{ds}=- \alpha_{s,\Lambda}^2 \; 
I_{s,\Lambda}^{(2)}(p,p')
\end{equation}
\noindent
where
\begin{eqnarray}
I_{s,\Lambda}^{(2)}(p,p')&=&\frac{1}{(2\pi)^2}\; \int_{0}^{\Lambda}dk \; k \; 
(2 k^2-p^2-p'^2)\; e^{-2s[p^2 p'^2+k^4-k^2(p^2+p'^2)]}\nonumber\\
&=&\frac{1}{(2\pi)^2}\; \frac{e^{-2s \; p^2 p'^2}}{4s} \; .
\end{eqnarray}
\noindent
The resulting second-order running coupling and irrelevant operator are given 
respectively by
\begin{equation}
\alpha_{s,\Lambda,2}=\frac{\alpha_{\Lambda}}{1-\frac{\alpha_{\Lambda}}{8\pi}\; 
\left[\gamma+{\rm ln}\left(2s \Lambda^4 \right)-{\rm Ei}\left(-2s \Lambda^4 
\right)\right]}
\end{equation}
\noindent
and
\begin{eqnarray}
F^{(2)}_{s,\Lambda}(p,p')&=&  \frac{1}{4}\left[\gamma+{\rm ln}(2s \; p^2 \; 
p'^2)-{\rm Ei}(-2s \; p^2 \; p'^2)\right]\nonumber\\
&+& \frac{1}{4}\left[\gamma+{\rm ln}(2s \Lambda^4)-{\rm Ei}(-2s 
\Lambda^4)\right]\nonumber\\
&-&\frac{1}{4}\left[\gamma+{\rm ln}\left(s \left[(p^2-\Lambda^2)^2 
+(p'^2-\Lambda^2)^2-(p^2-p'^2)^2\right]\right)\right.\nonumber\\
&&\left. \; \; \; \; \; -{\rm Ei}\left(-s \left[(p^2-\Lambda^2)^2 
+(p'^2-\Lambda^2)^2-(p^2-p'^2)^2\right]\right)\right]\; .
\end{eqnarray}

At third-order we have
\begin{eqnarray}
-\frac{1}{2 \pi}\; \frac{d{\alpha}_{s,\Lambda}}{ds}&+&\frac{1}{(2\pi)^2}\; 
\alpha_{s,\Lambda}^{2}\; 
\frac{dF^{(2)}_{s,\Lambda}(p,p')}{ds}+\frac{2}{(2\pi)^2}\; \alpha_{s,\Lambda} 
\; \frac{d\alpha_{s,\Lambda}}{ds}\; F^{(2)}_{s,\Lambda}(p,p')\nonumber\\
&+&\frac{1}{(2\pi)^3}\; \alpha_{s,\Lambda}^{3} \; 
\frac{dF^{(3)}_{s,\Lambda}(p,p')}{ds}=- \alpha_{s,\Lambda}^2 \; 
I_{s,\Lambda}^{(2)}(p,p')+ \alpha_{s,\Lambda}^3 \; I_{s,\Lambda}^{(3)}(p,p') \; 
,
\end{eqnarray}
\noindent
where
\begin{eqnarray}
I_{s,\Lambda}^{(3)}(p,p')&=&\frac{1}{(2\pi)^3}\;\int_{0}^{\Lambda}dk \; k \; (2 
k^2-p^2-p'^2)\; e^{-2s[p^2 p'^2+k^4-k^2(p^2+p'^2)]} \nonumber\\
&\times& \left[F^{(2)}_{s}(p,k)+F^{(2)}_{s}(k,p')\right] \; .
\end{eqnarray}
\noindent
In this case, 
\begin{eqnarray}
F^{(2)}_{s,\Lambda}(0,k)&=& \frac{1}{4}\left[\gamma+{\rm ln}(2s \Lambda^4)-{\rm 
Ei}(-2s \Lambda^4)\right]\nonumber\\
&=& \frac{1}{4}\left[\gamma+{\rm ln}(2s \Lambda^4-2s k^2 \Lambda^2)-{\rm 
Ei}(-2s \Lambda^4+2sk^2 \Lambda^2)\right]\; ,
\end{eqnarray}
\noindent
and so
\begin{eqnarray}
I_{s,\Lambda}^{(3)}(0,0)&=&\frac{1}{(2\pi)^3}\; \frac{1}{4}\; 
\int_{0}^{\Lambda}dk  \; 4 k^3 e^{-2s k^4}\; \left(\left[\gamma+{\rm ln}(2s 
\Lambda^4)-{\rm Ei}(-2s \Lambda^4)\right]\right. \nonumber\\
&-&\left. \left[\gamma+{\rm ln}(2s \Lambda^4-2s k^2 \Lambda^2)-{\rm Ei}(-2s 
\Lambda^4+2sk^2 \Lambda^2)\right]\right)\; .
\end{eqnarray}
\noindent
Since $I_{s,\Lambda}^{(3)}(0,0) \neq 0$ the term proportional to 
$\alpha_{s,\Lambda}^3$ in Eq. (\ref{alp32}) does not vanish. To obtain 
$\alpha_{s,\Lambda,3}$ we evaluate $I_{s,\Lambda}^{(3)}(0,0)$ and solve 
Eq. (\ref{alp32}) numerically.

The equation for $F^{(3)}_{s,\Lambda}(p,p')$ is given by 
\begin{equation}
\frac{1}{(2\pi)^3}\frac{dF^{(3)}_{s,\Lambda}(p,p')}{ds}=-\frac{1}{\pi} \; 
I_{s,\Lambda}^{(2)}(0,0)\; F^{(2)}_{s,\lambda}(p,p')+ I_{s,\Lambda}^{(3)}(p,p') 
\; .
\end{equation}
\noindent
To obtain $F^{(3)}_{s,\Lambda}(p,p')$ the integrals over $k$ and $s$ must be 
evaluated numerically. In the limit $s\Lambda^4 \rightarrow \infty$ with $s$ 
fixed at some non-zero value
\begin{eqnarray}
&&F^{(2)}_{s,\Lambda}(p,p')\rightarrow \frac{1}{4}\left[\gamma+{\rm ln}(2s \; 
p^2 \; p'^2)-{\rm Ei}(-2s \; p^2 \; p'^2)\right] \; , \\
&&I_{s,\Lambda}^{(3)}(0,0) \rightarrow 0 \; ,
\end{eqnarray}
\noindent
and $\alpha_{s,\Lambda}$ becomes indeterminate, requiring the coupling to be 
fixed at some $s_0 \neq 0$. In this way, we recover the result of the previous 
prescription.

Although less trivial, this prescription allows a more transparent error 
analysis. We can also extend the calculation to larger values of the similarity 
cutoff, $\lambda$, and analyze the errors in a different scaling regime. In 
Fig. 5 we show the errors in the binding-energy  obtained using the following 
approximations for the potential with $\Lambda=50$:
\vspace{0.5cm}

\noindent
(a) marginal operator with coupling ($\alpha_{\lambda_0}$),
\begin{equation}
V_{\lambda,\Lambda}(p,p')=-\frac{\alpha_{\Lambda}}{2\pi}\; 
e^{-\frac{(p^2-p'^2)^2}{\lambda^4}} \; ;
\end{equation}

\noindent
(b) marginal operator with running coupling renormalized to second-order 
($\alpha_{\lambda,\Lambda,2}$),
\begin{equation}
V_{\lambda,\Lambda}(p,p')=-\frac{\alpha_{\lambda,\Lambda,2}}{2\pi}\; 
e^{-\frac{(p^2-p'^2)^2}{\lambda^4}} \; ;
\end{equation}

\noindent
(c) marginal operator plus second-order irrelevant operator with running 
coupling renormalized to second-order ($\alpha_{\lambda,\Lambda,2}, 
F^{(2)}_{\lambda}$),
\begin{equation}
V_{\lambda,\Lambda}(p,p')=\left[-\frac{\alpha_{\lambda,\Lambda,2}}{2\pi}+\left(
\frac{\alpha_{\lambda,\Lambda,2}}{2\pi}\right)^2 \; 
F^{(2)}_{\lambda,\Lambda}(p,p')\right] \; e^{-\frac{(p^2-p'^2)^2}{\lambda^4}} 
\; ;
\end{equation}

\noindent
(d) marginal operator with running coupling renormalized to third-order  
($\alpha_{\lambda,\Lambda,3}$),
\begin{equation}
V_{\lambda,\Lambda}(p,p')=-\frac{\alpha_{\lambda,\Lambda,3}}{2\pi}\; 
e^{-\frac{(p^2-p'^2)^2}{\lambda^4}} \; ;
\end{equation}

\noindent
(e) marginal operator plus second-order irrelevant operator with running 
coupling renormalized to third-order $\alpha_{\lambda,\Lambda,3}, 
F^{(2)}_{\lambda}$),
\begin{equation}
V_{\lambda,\Lambda}(p,p')=\left[-\frac{\alpha_{\lambda,\Lambda,3}}{2\pi} 
+\left(\frac{\alpha_{\lambda,\Lambda,3}}{2\pi}\right)^2 \; 
F^{(2)}_{\lambda,\Lambda}(p,p')\right] \; e^{-\frac{(p^2-p'^2)^2}{\lambda^4}}  
\; .
\end{equation}
\vspace{0.5cm}

\begin{figure}
\centerline{\epsffile{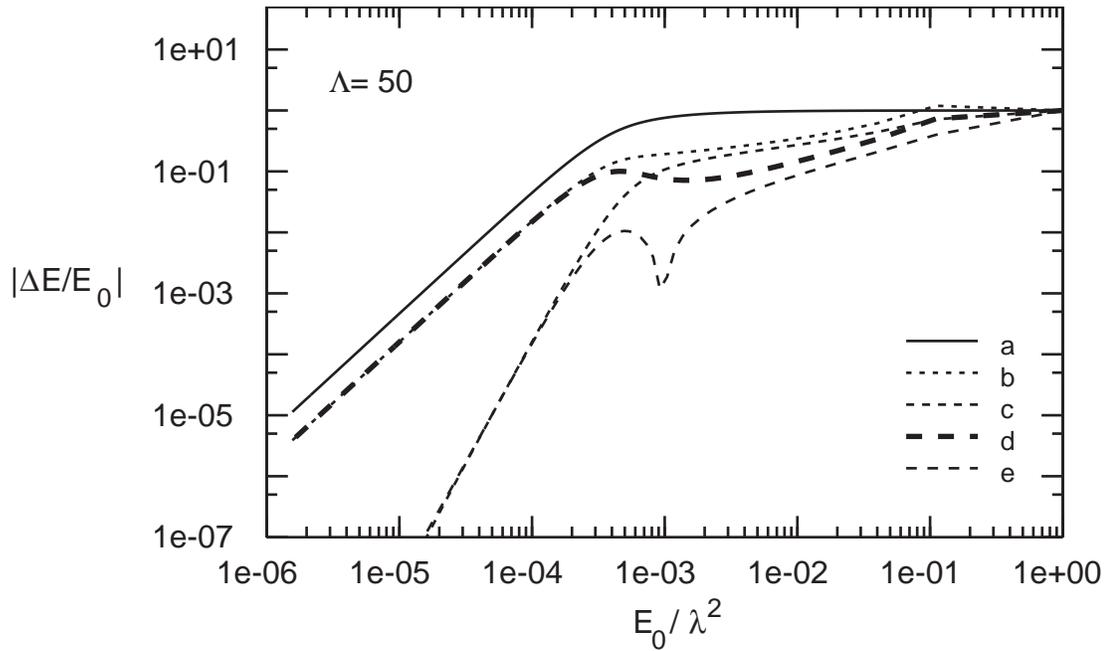}}
\vspace{0.5cm}
\caption{The SRG errors in the binding energy for the two-dimensional 
delta-function potential using various approximations for the similarity 
hamiltonian. The exact theory is fixed by regulating the ``bare hamiltonian'' 
using a sharp momentum cutoff, $\Lambda$, and letting the bare coupling depend 
on $\Lambda$ such that the binding energy is fixed. We use $\Lambda=50$ and 
$E_0=1$.}
\end{figure}

There are clearly two distinct scaling regions when an additional large
momentum cutoff $\Lambda$ is placed on the initial matrix and a similarity
cutoff is then applied. When the similarity cutoff is larger than
$\Lambda$, we see power-law improvement resulting from the addition of
irrelevant operators. Curves (a), (b), and (d) all have the same slope. None
of these hamiltonians contains irrelevant operators, but the marginal coupling
differs in each. All results become exact as the similarity cutoff goes to
infinity, and these curves are close to one another because the coupling runs
little in this region. Curves (c) and (e) show that there is a power-law
improvement when irrelevant operators are added, and that once again when the
similarity cutoff is larger than $\Lambda$ an improvement in the running
coupling makes little difference. Even though the coupling in front of this
operator is approximated by the first term in an expansion in powers of the
running coupling, the coupling is sufficiently small that this approximation
works well and the operator eliminates most of the leading power-law error in
curves (a), (b), and (d).

When the similarity cutoff become smaller than $\Lambda$ we see a crossover to
a more complicated scaling regime that resembles the SRG scaling discussed
above.  The error displayed by curve (a) approaches 100\%, while the running
coupling introduced in curve (b) reduces the error to an inverse logarithm.
Improving the running coupling in curve (d) further reduces the error, and we
see that curve (c) crosses curve (d) at a point where improving the running
coupling becomes more important than adding irrelevant operators. As above,
the best results require us to both improve the running coupling by adding
third-order corrections and add the second-order irrelevant operators. In no
case do we achieve power-law improvement, because as we have discussed there
are always residual inverse logarithmic errors. Had we fit the running
coupling to data, as we would do in a realistic calculation, we would obtain
power-law improvement and the residual error would be proportional to an
inverse power of the cutoff times an inverse power of the logarithm of the
cutoff. 

In Fig. 6 we show the running coupling at 2nd and 3rd order. Although the 3rd
order corrections are small for all $\lambda$ and vanish when $\Lambda
\rightarrow \infty$, the improvement resulting from this correction in Fig. 5
is significant.

\begin{figure}
\centerline{\epsffile{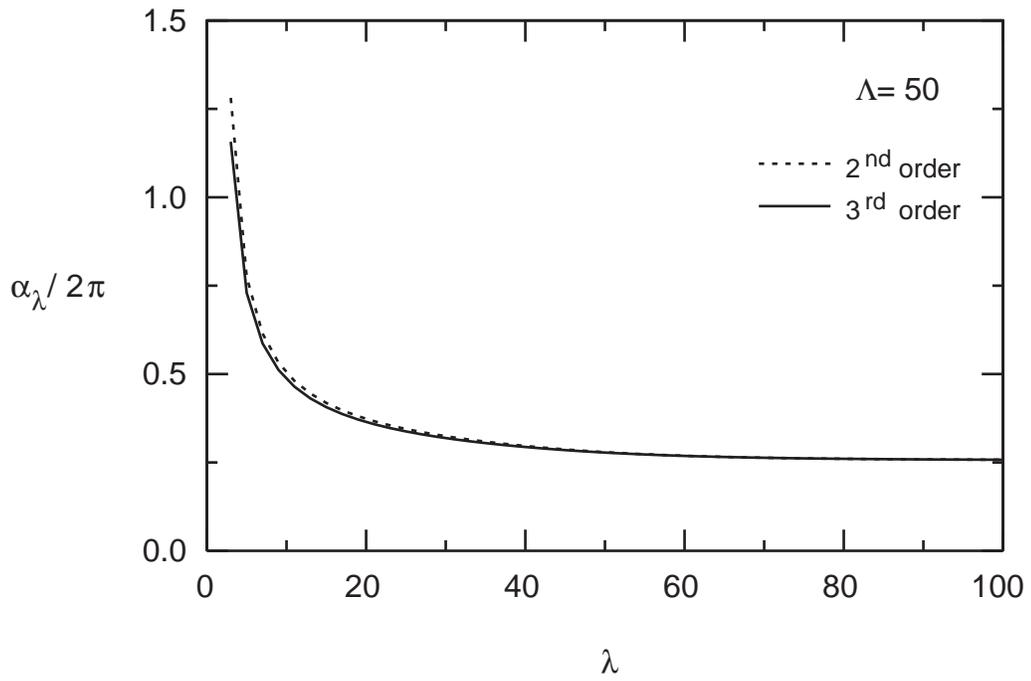}}
\vspace{0.5cm}
\caption{The SRG running coupling for the two-dimensional delta-function 
potential renormalized to second and third-order obtained with 
$\alpha_{\lambda_0=\infty}\rightarrow \alpha_{\Lambda}=4\pi/{\rm 
ln}\left(1+\frac{\Lambda^2}{E_0}\right)$.}
\end{figure}

The scaling behavior with a large momentum cutoff $\Lambda$ in place is
complicated, but it is fairly straightforward to understand it and to find a
sequence of approximations that systematically improve the non-perturbative
results. The calculations become increasingly complicated, but at each order
one must improve the running coupling, or fit it to data, and add higher order
irrelevant operators. In a field theory we need to let $\Lambda \rightarrow
\infty$ and study the scaling behavior of the theory in the regime where
$\lambda \ll \Lambda$. Although we do not display a compete set of figures, in
Fig. 7 we show what happens to the running coupling as $\Lambda$ is
increased, with the bound state energy fixed at one.

\begin{figure}
\centerline{\epsffile{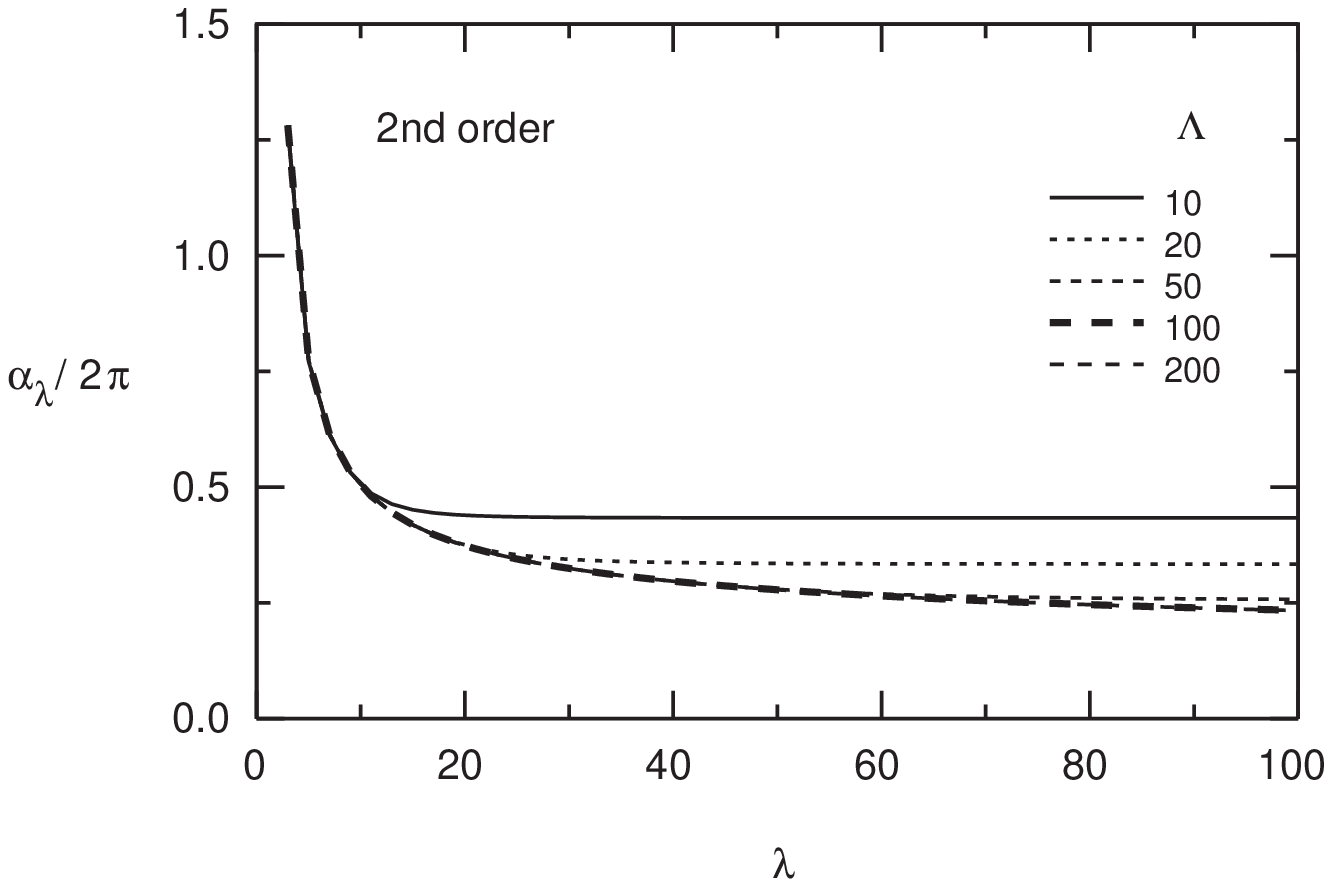}}
\vspace{0.5cm}
\caption{The SRG running coupling for the two-dimensional delta-function 
potential renormalized to second-order obtained with 
$\alpha_{\lambda_0=\infty}\rightarrow \alpha_{\Lambda}=4\pi/{\rm 
ln}\left(1+\frac{\Lambda^2}{E_0}\right)$.}
\end{figure}

As is evident in the exact solution, as $\Lambda$ increases the coupling
decreases. When $\lambda \gg \Lambda$, the coupling runs slowly and stays near
its asymptotic value. As $\lambda$ approaches $\Lambda$ the coupling begins to
run noticeably, and when $\lambda$ becomes much less than $\Lambda$ the
coupling approaches a universal curve that is insensitive to its asymptotic
value. Plots of the error in the binding energy for various approximations and
different values of $\Lambda$ closely resemble Fig. 11, with two scaling
regimes whose boundary is $\lambda = \Lambda$.

We close this section by reminding the reader that in all of these
calculations there is only one free parameter. In a realistic calculation we
would fit this parameter to a binding energy and we would expect to see
residual errors in other observables that is inversely proportional to powers
of the cutoff and logarithms of the cutoff.
%

%

%

\section{Conclusions}

We have illustrated the similarity renormalization group method for producing
effective cutoff hamiltonians using the two-dimensional delta-function
potential. We have shown that the SRG with coupling coherence leads to errors
that scale as inverse powers of the cutoff and inverse logarithms of the
cutoff. The SRG with coupling coherence requires the same number of parameters
as the underlying `fundamental' theory, but the cost is exponentially
increasing algebraic complexity to remove errors that contain inverse powers of
logarithms of the cutoff.

%
\section{Acknowledgments}

We would like to acknowledge many useful discussions with Brent Allen,
Martina Brisudova, Dick Furnstahl, Stan G{\l}azek, Billy Jones, Roger Kylin,
Rick Mohr, Jim Steele, and Ken Wilson. This work was supported by National
Science Foundation grant PHY-9800964, and S.S. was supported by a CNPq-Brazil
fellowship (proc. 204790/88-3).
%

%

\vspace{2cm}

\centerline{\LARGE \bf References}
\def\refname{}

\end{document}